\documentclass[graybox,vecphys]{svmult}

\usepackage{mathptmx}       
\usepackage{helvet}         
\usepackage{courier}        
\usepackage{type1cm}        
\usepackage{makeidx}         
\usepackage{graphicx}        
\usepackage{multicol}        
\usepackage[bottom]{footmisc}

\usepackage{epstopdf}
\usepackage{textcomp,amssymb,wasysym}

\newcommand{\mT}{\, \mathrm{mT}}
\newcommand{\mum}{\, \mbox{\textmu}\mathrm{m}}
\newcommand{\nm}{\, \mathrm{nm}}
\newcommand{\fs}{\, \mathrm{fs}}
\newcommand{\ps}{\, \mathrm{ps}}
\newcommand{\ns}{\, \mathrm{ns}}
\newcommand{\GHz}{\, \mathrm{GHz}}
\newcommand{\Hext}{\ensuremath{H_\mathrm{ext}}}
\newcommand{\muHext}{\ensuremath{\mu_0 H_\mathrm{ext}}}
\newcommand{\kDE}{\ensuremath{k_\mathrm{DE}}}
\newcommand{\MS}{\ensuremath{M_\mathrm{S}}}

\makeindex             

\begin{document}

\title*{Photo-magnonics}
\author{Benjamin Lenk, Fabian Garbs, Henning Ulrichs, Nils Abeling, and Markus M\"unzenberg}
\institute{Benjamin Lenk \at I. Institute of Physics, Georg-August-University of G\"ottingen \email{blenk@gwdg.de}}

\maketitle

\abstract{In the framework of magnonics all-optical femtosecond laser experiments are used to study spin waves and their relaxation paths. Magnonic crystal structures based on antidots allow the control over the spin-wave modes. In these two-dimensional magnetic metamaterials with periodicities in the wave-length range of dipolar spin waves the spin-wave bands and dispersion are modified.
Hence, a specific selection of spin-wave modes excited by laser pulses is possible.
Different to photonics, the modes depend strongly on the strength of the magneto-static potential at around each antidot site~--~the dipolar field.
While this may lead to a mode localization, also for filling fractions around or below 10\%, Bloch states are found in low damping ferromagnetic metals.
In this chapter, an overview of these mechanisms is given and the connection to spin-wave band spectra calculated from an analytical model is established. Namely, the plane-wave method yields flattened bands as well as band gaps at the antidot lattice Brillouin zone boundary.}

\section{Introduction}\label{sec:intro}
Being a quickly evolving research field, magnonics and magnonic materials have been investigated with different approaches~\cite{Lenk2011,Kruglyak2010}. These can be in the time or frequency domain, as well as with very high spatial resolution down to nanometers or averaging over wider areas, then giving more general information on the system in question.
For example, the spin-wave dispersion~$\omega(k)$ can precisely be measured using a vector network analyzer~(VNA)~\cite{Podbielski2006,NeusserPRL2010} or Brillouin light scattering~(BLS) setup~\cite{Tacchi2011,Sandweg2010}. In such experiments, the selection of the excitation frequency or wave vector~$k$, respectively, enables a detailed study of the magnetic mode spectrum in Fourier space.

In contrast, all-optical pump-probe techniques are neither $k$- nor frequency-selective. Instead, a broad continuum of spin-wave modes is populated on ultrafast time scales in the femto- to picosecond range by absorption of an intense laser pulse~\cite{marija07}. The heat-induced disorder can be modeled by high-$k$ spin-wave modes that subsequently relax into energetically lower-lying states.
A population of respective spin-wave modes leads to a spatial distribution of energy by spin-wave propagation away from the spot of (optical) excitation~\cite{Lenk2010}.
The underlying processes for excitation, relaxation, and propagation span a large range of interaction energies (i.e.\ time scales). Connected transient dynamics have an effect on the dielectric tensor and thus can be optically investigated.
Herein lies the large potential of (laser-) pulsed experiments: femto-, pico- and nanosecond characteristics can be resolved in a fast and non-destructive manner.

Concerning optically excited spin waves, these condensate-like modes are intrinsically dependent on the matrix material properties. However, the reverse process is also possible: In a magnonic structure, i.e.\ a spin-wave metamaterial, the material properties can be tailored to produce spin-wave modes whose characteristic properties are decoupled from the ferromagnetic matrix.
In the present chapter of this book, we shall restrict the discussion to two-dimensional systems. First, in section~\ref{subsec:conti} the continuous thin film case will be briefly described, while section~\ref{sec:samples} introduces all-optical experiments are.
The final sections~\ref{sec:Bloch-like-modes}~-~\ref{sec:nickel} are dedicated to structured media for which the manipulation of the spin-wave spectrum will be exemplified. Finally, an outlook on possible devices based on spin-wave computing is presented.

\subsection{Spin-wave modes in a thin ferromagnetic film}\label{subsec:conti}
For the case of a continuous film of thickness~$t$, the Landau-Lifshitz-Gilbert\index{Landau-Lifshitz-Gilbert equation} equation of motion can be solved analytically~\cite{de61,kalinikos86}.
On sufficiently large length scales, the exchange interaction may be neglected. The resulting spin-wave modes are of dipolar character and have theoretically been studied by Damon and Eshbach~(DE)\index{Damon-Eshbach!dispersion} in the 1960s. In the geometry of propagation perpendicular to the applied magnetic field (wave vector $\kDE \perp \Hext$) the dispersion\index{spin wave!dipolar} takes the form~\cite{de61}
\begin{equation}\label{eq:de}
\left(\frac{\omega_\mathrm{DE}}{\gamma\mu_0}\right)^2= H_x \left(H_x + M_\mathrm{S} - \frac{2K_z}{\mu_0M_\mathrm{S}}\right)+\frac{M_\mathrm{S}^2}{4}\Big(1-e^{-2 |k_\mathrm{DE}| t}\Big)\; ,
\end{equation}
where $\omega_\mathrm{DE}$ is the spin-wave frequency, $M_\mathrm{S}$ is the material's saturation magnetization\index{magnetization!saturation}, $H_x$ is the in-plane component of the external field\index{magnetic field!applied}, and $K_z$ accounts for an effective out-of-plane anisotropy\index{anisotropy!effective parameter}.
Respective wavelengths~$\lambda_\mathrm{DE} =2\pi / \kDE$ are in the micron range. On much smaller length scales, the exchange interaction has to be considered, while the dipolar interaction can then be neglected.
The dispersion for exchange-dominated spin waves reads
\begin{equation}\label{eq:pssw}
\left(\frac{\omega_\perp}{\gamma\mu_0}\right)^2=\left(H_x + \frac{2A}{M_\mathrm{S}}\, k_\perp^2\right) \left(H_x + M_\mathrm{S} - \frac{2K_z}{\mu_0M_\mathrm{S}} + \frac{2A}{M_\mathrm{S}}\, k_\perp^2\right)\; ,
\end{equation}
where $A$ is the exchange constant\index{exchange interaction!exchange constant}. In equation~(\ref{eq:pssw}), it has already been considered that respective spin-wave lengths are only relevant in the direction perpendicular to the film plane. Standing spin waves of the order~$n$ can be excited where the wave vector is then quantized according to $k_\perp=n\pi / t$~--~hence the naming perpendicular standing spin waves~(PSSW)\index{spin wave!perpendicular standing spin wave}.
The geometric confinement of thin films with thickness in the nanometer range implies the possibility to independently study exchange and dipolar spin waves. In the lateral directions, the exchange interaction does not play a significant role, whereas in the direction perpendicular to the film plane it determines the spin waves potentially excited.
From equations~(\ref{eq:de}) and~(\ref{eq:pssw}), another precessional mode can be deduced: An in-phase precession of all spins, i.e.,\ the uniform mode of ferromagnetic resonance\index{ferromagnetic resonance} (Kittel mode\index{Kittel mode}~\cite{hk51}), will have the dispersion~$\omega_\mathrm{K} = \omega_\mathrm{DE}|_{k=0} = \omega_\perp|_{k=0}$.
Equations~(\ref{eq:de}) and~(\ref{eq:pssw}) together with the Kittel mode, constitute the basic possibilities of magnetic excitations in thin ferromagnetic films. In the following section~\ref{sec:samples} a brief example of a continuous film investigated all-optically is given.

\section{Samples and experiments}\label{sec:samples}
Ferromagnetic thin films made of nickel\index{nickel} or cobalt-iron-boron~(CoFeB)\index{CoFeB} are produced by electron-beam evaporation or magnetron sputtering~\cite{Eilers2009}. The magnonic metamaterials' properties cannot be simply derived from the independent properties of continuous film and (periodically arranged) constituents.
Instead, effects emerge that are intrinsic to the newly created material, i.e.,\ the observed collective effects themselves define the respective system as being a metamaterial. Consequently, different periodic structures on a continuous film with otherwise identical properties yield different magnonic materials.
\begin{figure}[t]
\includegraphics[width=\linewidth]{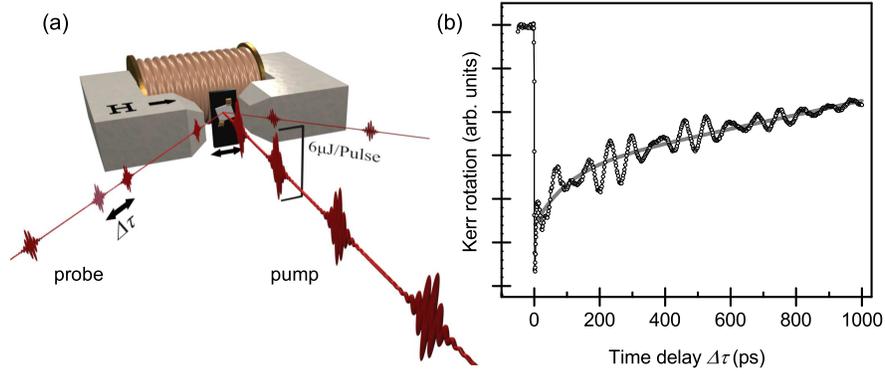}
\caption{A schematic of the time-resolved all-optical experiments is given in~(a). Pump pulses with a high intensity excite the magnetization dynamics and probe pulses that are time-delayed by $\Delta\tau$ measure the pump-induced change in the Kerr rotation. A reference data set recorded on a $50\nm$ CoFeB film at $\muHext=150\mT$ is shown in~(b). The solid gray line represents an exponential fit to the background which is subtracted prior to further analysis.}
\label{fig:trmoke}
\end{figure}

In photo-magnonics\index{photo-magnonics}, all-optical experiments are performed using ultrashort laser pulses to both excite and detect spin-waves. A schematic drawing of the experiment is given in figure~\ref{fig:trmoke}(a).
It includes the intense optical pump pulses with a central wavelength of $800\nm$ and a temporal duration of $40\fs$. Probe pulses with 5\% of the intensity at a variable time delay~$\Delta\tau$ yield the electron or magnetization dynamics~\cite{Marija06JAP}.
The excitation by the pump pulses takes place within picoseconds, including the ultrafast demagnetization ($\approx 100\fs$)~\cite{JakobPRL2010} and the relaxation of the highly damped high-$k$ spin-wave modes to longer-lived ones ($\approx 20\ps$)~\cite{marija07}.
The respective processes can be experimentally accessed making use of the time-resolved magneto-optical Kerr effect~(TRMOKE)\index{TRMOKE}.
Shown in Fig.~\ref{fig:trmoke}(b) are the precessional modes observed with probe pulses that are time-delayed by $\Delta\tau$ with respect to the pump pulses.
The observed modes have frequencies in the GHz range, thus oscillate on timescales up to nanoseconds. In consequence, the non-selectivity in terms of wave vector or frequency of the optical pump can be compared to a delta-like excitation of a classical oscillator which then will swing with its intrinsic eigen-frequency.

\subsection{Thin-film magnetization dynamics}
\begin{figure}[t]%
\centering%
\includegraphics[width=\linewidth]{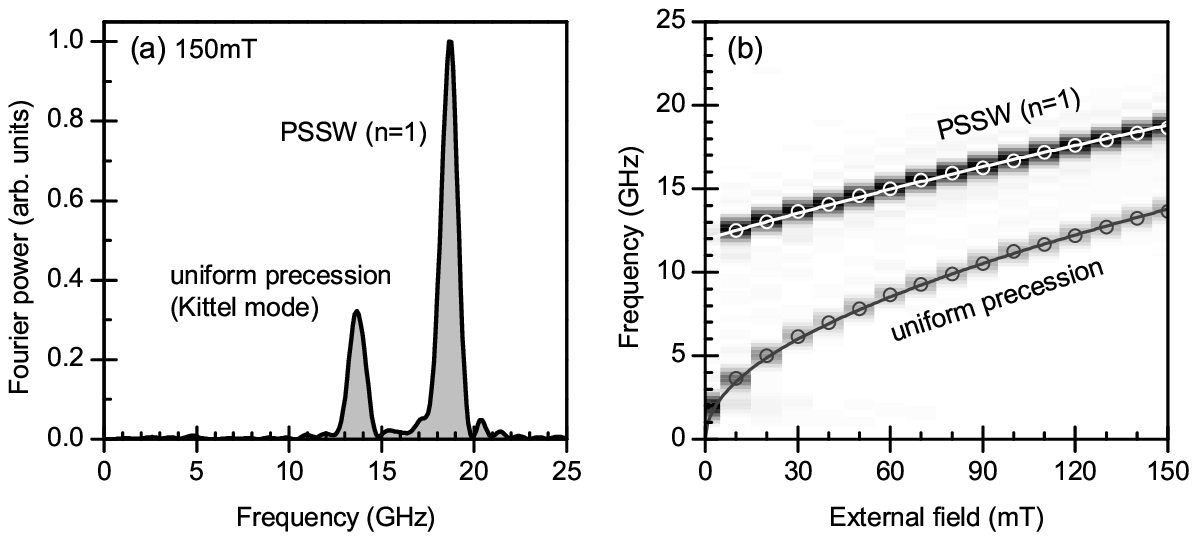}
\caption{Fourier analysis of the time-resolved magnetization dynamics of a continuous CoFeB film. In~(a) the Fourier spectrum\index{Fourier spectrum} of the single measurement already shown in Fig.~\ref{fig:trmoke} is displayed ($\muHext=150\mT$). Two precessional modes are observed the frequency of which is dependent on the external magnetic field~(b)\index{spin wave!perpendicular standing spin wave}. Circles represent experimentally determined peak positions, the gray and white solid lines are fits of equation~(\ref{eq:pssw}) for the case of $k_\perp=0$ and $k_\perp \neq 0$, respectively.}%
\label{fig:conti}%
\end{figure}

The reference data set in Fig.~\ref{fig:trmoke}(b) was recorded on a continuous CoFeB film of thickness~$t=50\,\mathrm{nm}$ for $\muHext=150\mT$. It shows the fast demagnetization~\cite{beau96} upon absorption of the pump pulse (time delay~$\Delta\tau=0$) and the subsequent magnetization dynamics which are a superposition of at least two precessional modes~\cite{Marija06JAP}.
The particular trace $M(\Delta\tau)$ from Fig.~\ref{fig:trmoke}(b) is further analyzed in Fig.~\ref{fig:conti}(a) by a Fourier transformation. The two modes' frequencies can then be determined. By changing the external field, these frequencies shift and the resulting experimental dispersions $\omega_i(\Hext)$ can be fitted by Eq.~(\ref{eq:pssw}) as seen in figure~\ref{fig:conti}(b).
All data presented in this chapter are plotted in a gray scale, where the (meta)material's response (i.e.,\ the excited spin-wave mode's Fourier power) is indicated by a black peak as a function of externally applied magnetic field and frequency in Fourier space. The maximum time delay between pump and probe pulses of $1\ns$ consequently limits the resolution to be $(1\ns)^{-1}= 1\GHz$, in the plots represented by the peak line width.

On the reference film, two modes of precession are observed. These are the uniform Kittel mode (white circles and line) as well as the first order perpendicular standing spin wave (PSSW, gray circles and line).
Due to the intriguingly low Gilbert damping\index{damping!FMR} in ferromagnetic CoFeB, standing spin waves up to high orders~$n \leq 5$ are observed in continuous films. While the spin-wave modes carry information on the structural properties of the film material~\cite{Lenk2010}, we will in the following focus on periodically structured materials.

\section{Bloch-like modes in CoFeB antidot lattices}\label{sec:Bloch-like-modes}
\begin{figure}[t]%
\includegraphics[width=\linewidth]{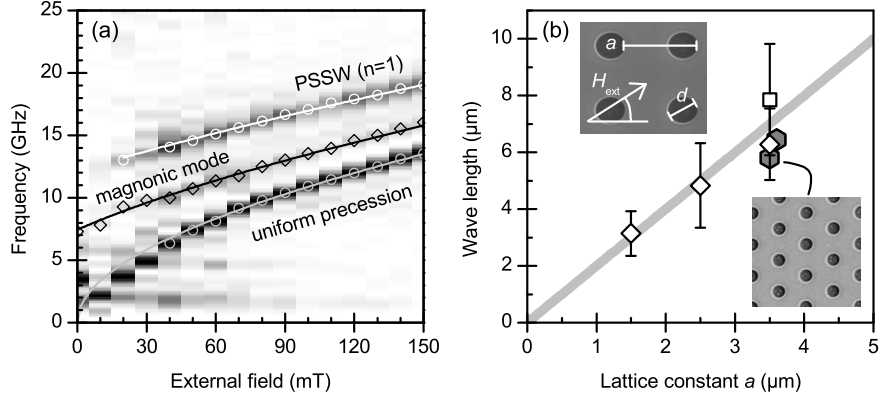}%
\caption{Magnetization dynamics on structured CoFeB films. The Fourier spectrum in~(a) shows an optically excited magnonic mode (black diamonds) additional to the modes also observed on a continuous film (gray and white circles and lines, see Fig.~\ref{fig:conti}). The solid black line is a fit of Eq.~(\ref{eq:de}) which yields the magnonic mode's wave vector~$k$ further detailed in~(b): A change in the antidots' periodicity~$a$ shifts the Bloch-like resonance according to $k = \pi/a$ (gray line). Included are points for propagation along different directions with respect to \Hext, namely $90^\circ$~($\square$) and $45^\circ$~($\Diamond$) in a square lattice, as well as $30^\circ$ and $60^\circ$ in a hexagonal lattice~($\varhexagon$). The insets of~(b) show SEM images of a square and hexagonal antidot-lattice unit cell, respectively.}%
\label{fig:struct-tune-k}%
\end{figure}

As described earlier, a metamaterial cannot simply be defined by a relation between the wavelengths and the parameters of the imprinted periodic structure. Instead, it is the emerging effects that make the name metamaterial necessary. For example, in a structured CoFeB film a new~--~magnonic~--~spin-wave mode is observed~\cite{Ulrichs2010}.
An SEM image of the two-dimensional square lattice of antidots in an otherwise continuous film is included in the inset of figure~\ref{fig:struct-tune-k}(b). It depicts the antidot diameter~$d$ and the lattice parameter~$a$.
Compared to the reference data set from Fig.~\ref{fig:conti}, the data recorded on a structured film which are plotted in Fig.~\ref{fig:struct-tune-k}(a) reveal another precessional mode not previously observed (black diamonds and line).
With the Damon-Eshbach dispersion~$\omega_\mathrm{DE}(\Hext)$ the magnonic character is verified: Fitting Eq.~(\ref{eq:de}) to the experimental peak positions yields (as the only fitting parameter) the wave vector to be $\kDE = \pi/a$.
From that value and the observed frequency, the spin-wave propagation length can be approximated via the phase velocity: Together with the damping time constant determined in the TRMOKE data, one calculates length scales of about $100\mum$ which equal approximately 30 magnonic unit cells.
This leads to an instructive picture on Bloch-mode excitation in real space: the periodic modulation of the spin-wave potential landscape (i.e.,\ the effective internal magnetic field) imposes its periodicity as a condition on the spin waves propagating away from the spot of excitation.
In other words, prerequisites for the observation of spin-wave Bloch states are the low damping in CoFeB and the resulting large propagation length of the spin waves.
By changing the periodicity~$a$ of the antidot lattice per definition, a new metamaterial is created, in the sense that the magnonic properties are changed. Namely, the DE wave vector can be tuned according to the above-stated relation over a range of $1.5\mum  \leq a \leq 3.5\mum$.
For the measurements in Fig.~\ref{fig:struct-tune-k}(b), the filling fraction~$f = \pi d^2 / (4 a^2) = 12\%$\index{filling fraction} has been held constant. For small antidot separations, an additional effect comes into play: besides the Bloch-like extended modes, also localized ones are excited. These will be further detailed in section~\ref{sec:nickel}.

\subsection{Effects of antidot-lattice symmetry}
Included in Fig.~\ref{fig:struct-tune-k}(b) are the experimental results obtained in configurations deviating from $\kDE\perp\Hext$. For example, the antidot lattice can be rotated by $45^\circ$ around the film normal. In that case, the Damon-Eshbach dispersion~(\ref{eq:de}) cannot be fitted to satisfyingly represent the experimentally observed magnonic mode dispersion.
This deviation can be accounted for by choosing the angle between propagating surface waves and $\Hext$ to match $45^\circ$. The dispersion then reads~\cite{kalinikos86}
\begin{eqnarray}\label{eq:de-45}%
\left(\frac{\omega_{45}}{\gamma\mu_0}\right)^2= H_x^2 &+& \frac{H_x \MS}{2}\left[1 + \frac{1}{k_{45}t}\left(1-e^{-k_{45}t}\right)\right]+ \nonumber\\
&+& \frac{\MS^2}{2k_{45}t}\left(1-e^{-k_{45}t}\right) \left[1- \frac{1}{k_{45}t}\left(1-e^{-k_{45} t}\right)\right]\; .
\end{eqnarray}%
Respective fitting results for $k_\mathrm{45}$ are included in Fig.~\ref{fig:struct-tune-k}(b) as white diamonds. One finds the relation $k_{45} = \pi/a$ to be satisfied. This means that by rotation of the sample the propagation direction of the Bloch-like surface waves is not changed~--~they still preferably propagate along the nearest-neighbor directions of the antidot lattice.

This remains true for lattices with other than square symmetry. Also in films structured with hexagonal lattices magnonic spin-wave modes are observed. Here, the application of the external field along a high-symmetry axis and assuming a propagation direction in the nearest-neighbor direction (under $30^\circ$ and $60^\circ$, respectively) yields the wave vector~$k_\mathrm{hex} = \pi/a$ .
Respective data points are represented in Fig.~\ref{fig:struct-tune-k}(b) by the large filled hexagons.

The population of the magnonic modes suggests the creation of flattened bands that increase the spin-wave density of states~(DOS)\index{spin wave!density of states}.
For spin-waves, the situation is analogous to electrons in a crystal, where a periodic potential is created by the atoms. More precisely, the spin-wave spectrum is modified by the antidots such that band gaps at the Brillouin zone boundary\index{Brillouin zone!boundary} are introduced.
Simultaneously, a flattening of bands takes place which in turn leads to an increase of the density of states. Keeping in mind the condensation-like excitation of precessional modes after optical excitation, this increase of the spin-wave DOS can be considered  the reason for the observation of the Bloch-like modes.
The broad-band excitation by the pump pulses provides the range of spin waves initially necessary to populate the modes discussed here.
The close analogy to electronic crystals will be extended in section~\ref{sec:PWM} where a calculation of the spin-wave spectrum is performed via the plane-wave method.

\section{Spin-wave spectra from plane-wave calculations}\label{sec:PWM}
An analytical approach to theoretically access the creation of band gaps and the flattening of bands is via the plane-wave method~(PWM)\index{plane-wave method}. It has been developed by Puszkarski et al.\ and can be used to study the appearance of spin-wave band gaps in two-dimensional~\cite{Vasseur1996}, as well as three-dimensional systems~\cite{Krawczyk08}.
For the present case of a thin film with thickness~$t=50\nm$ and wavelengths~$\lambda$ in the micron range, the assumption $\lambda \gg t$ holds.
Thus, the lower modes' profiles can be assumed to be uniform in the $z$-direction, i.e.,\ perpendicular to the thin film. As opposed to the full theory in~\cite{kalinikos86}, the exchange interaction can then be neglected for sufficiently small~$k$ and the uniform-mode analysis applied~\cite{Hurben1995}.
The respective spin waves are confined in a periodic potential which is constructed by a periodic modulation\index{magnetization!periodic modulation} of the magnetization in terms of a Fourier synthesis:
\[
M_\mathrm{S}( \vec{r} )=\sum_{\vec{G}} M_\mathrm{S}( \vec{G} ) e^{i \vec{G} \vec{r} }\; ,
\]
where $\vec{G}$ is a two-dimensional vector of the reciprocal lattice. Using this formulation, the linearized Landau-Lifshitz-Gilbert equation can be solved. Namely, plane waves of a given wave vector are used to solve the eigenvalue problem stemming from the dynamic field and $M_\mathrm{S}( \vec{r} )$~\cite{Lenk2011}. The resulting frequency for each mode depends on the position and direction in the reciprocal space of the antidot lattice.
This ansatz is similar to the Bloch theorem applied to electrons and photons and yields the magnonic band structure of a periodic ferromagnetic system.
\begin{figure}[t]
\includegraphics[width=\linewidth]{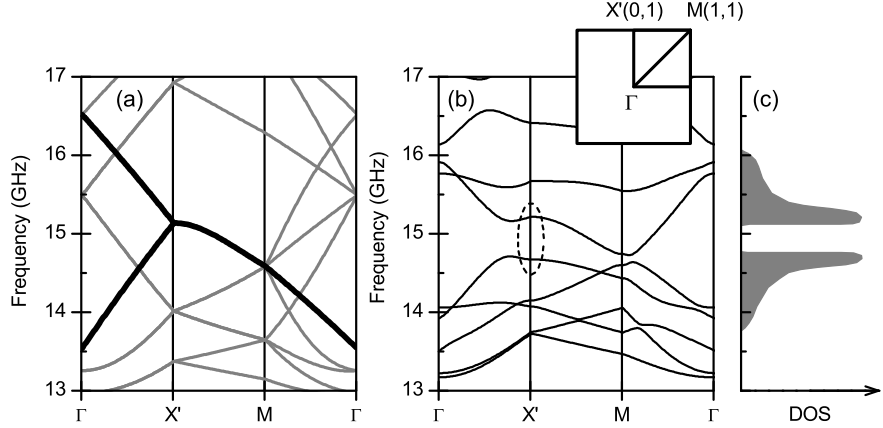}
\caption{Spin-wave band structures of (a)~a continuous CoFeB film and (b)~an antidot lattice with $a=3.5\mum$ and $d=1\mum$ in a CoFeB matrix ($\muHext = 130\mT$). In~(a), the first and second band are marked by the thicker black lines. In~(b), the band gap created by the periodic structure is highlighted by the dashed black ellipsis. In the inset, the first Brillouin zone\index{Brillouin zone!magnonic} of the square lattice is depicted. The reciprocal directions along~$X'$ correspond to $k \perp \Hext$ in real space. Schematically depicted in~(c) is the spin-wave density of states~(DOS)\index{spin wave!density of states} for the 1st and 2nd band in this direction, which is populated by the laser pump pulses.}
\label{fig:bands}
\end{figure}

Respective results are plotted in figure~\ref{fig:bands}. Therein, a calculation for a continuous CoFeB film is compared to a structured film\footnote{continuous means $d \to 0$ while $a=3.5\mum$ is held constant}.
In the continuous film, the folding of the spin-wave bands at the Brillouin zone boundary is introduced to illustrate the difference when compared to the structured film.
For the latter, at the point~$X'$ in reciprocal space, a band gap as marked by the dashed ellipsis opens up around a frequency of $15\GHz$. This corresponds well to the experimentally observed magnonic mode in Fig.~\ref{fig:struct-tune-k}(b).
Hence, there is a close relation between the flattening of the spin-wave band at the zone boundary and the experimental finding of the wave vector being equal to $\pi/a$.
The situation is further illustrated in Fig.~\ref{fig:bands}(c). Therein, the density of states as a function of the spin-wave frequency is schematically depicted for the 1st and 2nd band in the $\Gamma-X'$ direction. The peaks introduced by the periodic structure correspond to the experimentally observed spin-wave modes.

Details on the calculations and the application to different ferromagnets can be found in~\cite{Lenk2011}. Concerning size and position of the band gap, tuning is possible by the material parameters of the constituent materials~\cite{Krawczyk08}.

\section{Localized modes in nickel antidot lattices}\label{sec:nickel}
\begin{figure}[t]%
\sidecaption[t]%
\includegraphics[width=0.6\linewidth]{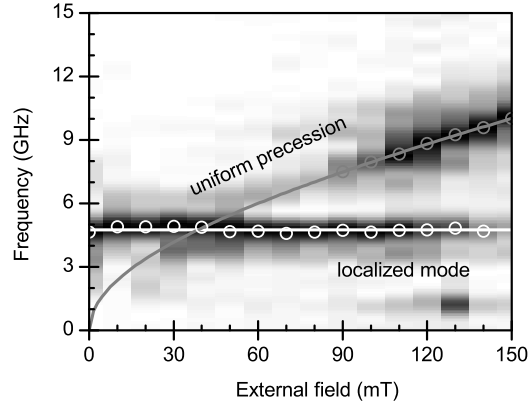}%
\caption{Magnetization dynamics on a structured nickel film with $a=3.5\mum$ and $d=1\mum$. The strong intrinsic damping leads to the observation of only a localized magnetic mode  (white circles and line). Its amplitude is largest around the antidot edges, and its frequency does not depend on the external field. Only at high $\muHext \geq 70\mT$, the uniform precession is excited by the pump pulses (gray circles and line).}%
\label{fig:nickel}%
\end{figure}

TRMOKE experiments on magnonic crystals with a nickel matrix do not show any Bloch modes as observed in CoFeB in section~\ref{sec:Bloch-like-modes}. Instead, a field-independent mode occurs as shown in Fig.~\ref{fig:nickel}. Its frequency is independent of \Hext\ and remains constant up to $\muHext = 150\mT$.
Moreover, its amplitude is larger than the one of the uniform mode. Only if the external field is large enough, the excitation efficiency for the Kittel mode is high.
This already points to an inhomogeneity of the internal field: If \Hext\ is sufficiently high, regions of constant internal field exist, where the Kittel mode can be populated.
Below the critical field, the internal field undergoes strong spatial changes and localization sites for spin waves are formed~\cite{PechanJAP2005}.
This is consistent with the experimental finding that the filling fraction~$f$ merely changes the amplitude of the localized mode, but does not alter its field-independence.

It has to be considered that the observed effects remain a combination of matrix material and superimposed periodic structure.
In ferromagnetic nickel, the intrinsic Gilbert damping is by a factor of 4 larger than in CoFeB. This leads to a strongly decreased propagation length of spin waves, which is not more than $10\mum$ in nickel (as opposed to $100\mum$ in CoFeB, see section~\ref{sec:Bloch-like-modes}).
Consequently, the spin waves excited by laser pulses do not propagate far enough to largely interact with the periodic structure of antidots.
However, also in CoFeB localized modes are observed, if the periodic distortion becomes too strong. For an antidot separation in the order of the internal field inhomogeneity, additional to the propagating magnonic mode a non-dispersive localized mode appears~\cite{Ulrichs2010}.

The frequency of the localized modes can be shifted by virtue of the antidots' separation. This points to a dipolar coupling of individual locations, i.e.,\ regions of strong field inhomogeneity. The same effect can be achieved by rotation of the external field with respect to the antidot lattice.
The dipolar coupling between neighboring localization sites changes and hence, the field-independent modes' frequency is shifted, see Fig.~27 in Ref.~\cite{Lenk2011}.

\section{Outlook: Magnonic control over spin waves}\label{sec:applications}
The different effects of a periodic antidot structure in two dimensions have been described. The respective metamaterial(s) illustrate the large potential of artificially structured media in general and magnonics in particular.
Namely, they allow the control over spin waves in that the two rivalling effects of localization and Bloch-like extension can be tuned. The selective excitation of magnonic states from a broader continuum of spin waves carries information on the spin-wave density of states.
In that respect, the applicability of basic concepts of solid state physics to magnonic crystals has been shown.

Relevant parameters that determine the metamaterials' properties are not only the constituting material, but very importantly the structural parameters of the antidot lattice inside.
If, for example, the intrinsic spin-wave propagation length is long enough, information may be carried across multiple unit cells in form of a Bloch-like spin wave. On the other hand, if the intrinsic damping is larger or the filling fraction is too high, localized edge modes are observed that may be used to collect information on the structures' internal field.
In combination with other approaches described in the present book, e.g.,\ different material combinations, the perspectives are promising: if taken beyond the proof of principle, spin-wave computation and spin-wave data transmission may be combined to provide a high data throughput and energy efficiency in such devices.
One example for a distinct modification of the magnonic periodic structure is a line defect that can function as a wave guide for spin-wave frequencies inside the magnonic gap region, to be functionalized in the future.

\begin{acknowledgement}
We would like to thank Jakob Walowski for the contribution of the experiment's schematic shown in Fig.~\ref{fig:trmoke}(a) and Georg Herink for careful proof-reading of the manuscript.
\end{acknowledgement}

\bibliographystyle{spphys}
\bibliography{magnonics-chapter-photo-magnonics}


\end{document}